\documentclass[english,aps,manuscript,superscriptaddress,prd,aps,preprint,showpacs]{revtex4}
\usepackage[latin9]{inputenc}
\usepackage{float}
\usepackage{amsmath}

\makeatletter

\DeclareTextSymbolDefault{\textquotedbl}{T1}
\providecommand{\tabularnewline}{\\}

\@ifundefined{textcolor}{}
{%
 \definecolor{BLACK}{gray}{0}
 \definecolor{WHITE}{gray}{1}
 \definecolor{RED}{rgb}{1,0,0}
 \definecolor{GREEN}{rgb}{0,1,0}
 \definecolor{BLUE}{rgb}{0,0,1}
 \definecolor{CYAN}{cmyk}{1,0,0,0}
 \definecolor{MAGENTA}{cmyk}{0,1,0,0}
 \definecolor{YELLOW}{cmyk}{0,0,1,0}
}

\usepackage{amsfonts}
\usepackage{slashed}

\usepackage{babel}

\makeatother

\usepackage{babel}
\begin{document}
\title{Radiative corrections and Lorentz violation}
\author{A. F. Ferrari}
\affiliation{Universidade Federal do ABC, Centro de Ciências Naturais e Humanas,
Rua Santa Adélia, 166, 09210-170, Santo André, SP, Brasil}
\email{alysson.ferrari@ufabc.edu.br}

\affiliation{Indiana University Center for Spacetime Symmetries, Indiana University,
Bloomington, Indiana 47405-7105}
\author{J. R. Nascimento}
\affiliation{Departamento de Física, Universidade Federal da Paraíba, Caixa Postal
5008, 58051-970, João Pessoa, Paraíba, Brazil}
\email{jroberto, petrov@fisica.ufpb.br}

\author{A. Yu. Petrov}
\affiliation{Departamento de Física, Universidade Federal da Paraíba, Caixa Postal
5008, 58051-970, João Pessoa, Paraíba, Brazil}
\email{jroberto, petrov@fisica.ufpb.br}

\begin{abstract}
Radiative corrections in Lorentz violating (LV) models have already
received a lot of attention in the literature in recent years, with
many instances where a LV operator in one sector of the Standard Model
Extension (SME) generates, via loop corrections, one of the LV coefficients
in the photon sector, which is probably the most understood and well
constrained part of the SME. In many of these works, however, the
now standard notation of the SME is not used, which can obscure the
comparison of different results, and their possible phenomenological
relevance. In this work, we fill this gap, trying to build up a more
general perspective on the topic, bringing many of the results to
the SME conventional notation and commenting on their possible phenomenological
relevance. We uncover one example where a result already presented
in the literature can be used to place a stronger bound on the temporal
component of the $b_{\mu}$ coefficient of the fermion sector of the
SME. 
\end{abstract}
\maketitle

\section{Introduction}

The idea that Lorentz symmetry might be violated by new physics at
the Planck scale is one of the motivations for the development of
the Standard Model Extension (SME)\,\citep{coll1,coll2} as an effective
field theory based on the internal symmetries and field content of
the Standard Model, incorporating a very general set of Lorentz violating
operators. In the minimal SME, power counting renormalizability is
enforced, so that only operators with mass dimensions four or less
are included, while the non-minimal extension of the SME includes
all operators with higher mass dimensions\,\citep{KostelMewes-EMHD,neutrinosHD,KosMew,Ding:2016lwt},
thus incorporating a huge set of new terms and presumably new physics.
Lorentz symmetry being one of the cornerstones of QFT as we know it,
Lorentz violation (LV) brings with it many interesting theoretical
questions; at the same time, a fruitful experimental programme has
used the SME framework to obtain new and improved tests of Lorentz
invariance from many experiments and astrophysical observations\,\citep{datatables,cpt2016}.

The source of LV in the SME is generally assumed to be new physics
at some high energy scale, for example spontaneous symmetry breaking
in a more fundamental theory such as string theory\,\citep{LVstrings},
in which case the constant tensors that couple to the LV operators
arise as vacuum expectation values of tensor fields in this theory.
One may also consider explicit breaking, which amounts to assume some
unknown mechanism generating LV at the fundamental level, and therefore
the LV background tensors are taken to assume, in principle, unspecified
non-zero values. In the non gravitational sector of the SME, the difference
between explicit and spontaneous LV breaking is not essential in principle,
but in the gravitational sector, one finds that explicit breaking
is in general incompatible with the usual geometric picture of general
relativity\,\citep{KosGra}. In this work, we will be mostly interested
in the non-gravitational sector, so LV can be assumed to be explicit
for simplicity.

As an example, Lorentz violation can be included in the standard Maxwell
theory, leading to the photon sector of the SME, which is probably
its most well studied part, being of utmost importance from the phenomenological
viewpoint since the most stringent constraints on Lorentz violation
are generally obtained by studying LV effects on photon propagation.
It is described by the Lagrangian density
\begin{align}
{\cal L}_{photon}= & -\frac{1}{4}F^{\mu\nu}F_{\mu\nu}+\frac{1}{2}\epsilon^{\kappa\lambda\mu\nu}A_{\lambda}(\hat{k}_{AF})_{\kappa}F_{\mu\nu}\nonumber \\
 & -\frac{1}{4}F_{\kappa\lambda}(\hat{k}_{F})^{\kappa\lambda\mu\nu}F_{\mu\nu}\thinspace,\label{eq:Lph}
\end{align}
where 
\begin{subequations}
\label{eq:genKAF}
\begin{align}
(\hat{k}_{AF})_{\kappa} & =\sum_{d\text{ odd}}(\hat{k}_{AF}^{\left(d\right)})_{\kappa}^{\hphantom{\kappa}\alpha_{1}\cdots\alpha_{(d-3)}}\partial_{\alpha_{1}}\cdots\partial_{\alpha_{\left(d-3\right)}}\thinspace,\\
(\hat{k}_{F})^{\kappa\lambda\mu\nu} & =\sum_{d\text{ even}}(\hat{k}_{F}^{(d)})^{\kappa\lambda\mu\nu\alpha_{1}\cdots\alpha_{(d-4)}}\partial_{\alpha_{1}}\cdots\partial_{\alpha_{\left(d-4\right)}}\thinspace,
\end{align}
\end{subequations}
$d\geq3$ being the dimension of the corresponding operator (for more
details see\,\citep{KostelMewes-EMHD}). The minimal photon sector
of the SME is obtained by the restriction to $d=3$ and $d=4$ for
the CPT-odd and CPT-even terms, corresponding to the original $k_{AF}$
and $k_{F}$ defined in\,\citep{coll2}. Some of the fundamental
references regarding this sector of the SME are presented by\,\citep{CFJ,Kostelecky:2001mb,KosteleckyMewes2002,Bailey:2004na},
and the most recent experimental limits can be found in\,\citep{datatables}.
It is noteworthy to recall that the $k_{AF}$ coefficient, and parts
of the $k_{F}$, induce birefringence in the vacuum, which leads to
very strong experimental constraints, from astrophysical observations:
components of $k_{AF}$ are constrained to the order $10^{-43}\text{GeV}$,
while the birefringent components of $k_{F}$ to the order of $10^{-37}$. 

From the theoretical viewpoint, the quantum impacts of the LV operators
in the SME are a matter that already received much attention. Lorentz
symmetry being one of the starting points of the usual approach to
quantum field theory, whether the full renormalization program can
be carried out consistently or not for the minimal SME, which is power
counting renormalizable, is an interesting question, which has been
positively answered for the QED\,\citep{KostPic}, electroweak\,\citep{Colladay:2009rb},
scalar and Yukawa\,\citep{Ferrero:2011yu} sectors. Interesting results
have also been obtained regarding the Källén-Lehman representation\,\citep{Potting:2011yj}
and the properties of asymptotic states\,\citep{Cambiaso:2014eba}
due to Lorentz violation, showing that much of the structure of QFT
survives the introduction of LV as done in the SME, but space still
remains for nontrivial modifications. Even from the phenomenological
point of view, the study of quantum corrections might be of interest,
since it can connect LV coefficients in different sectors of the framework,
thus making it possible to transfer bounds found in one sector to
the other. The general motivation can be explained by recalling the
first example of this mechanism\,\citep{coll2,Coleman:1998ti,Chung:1998jv,Andrianov:1998ay,JK},
where integration over a fermion loop, including a LV operator involving
an axial vector $b^{\mu}$, was shown to generate a finite correction
to the photon effective action, proportional to the well known Carroll-Field-Jackiw
(CFJ) term\,\citep{CFJ} (in the SME nomenclature, that means the
$b$ term in the QED fermion sector generating, by radiative correction,
the $k_{AF}$ coefficient in the photon sector). Strong experimental
bounds on the CFJ coefficient (which are still among the strongest
found in the literature so far\,\citep{datatables}) were already
derived in\,\citep{CFJ}, and theoretical consistency of the mechanism
presented in\,\citep{JK} would allow one to translate this bounds
to the original $b^{\mu}$ coefficient. However, from the start it
was recognized that the generated CFJ term was finite, yet ambiguous,
its value depending on the regularization scheme, thus being one more
example of \textquotedbl finite but undetermined\textquotedbl{} quantum
corrections\,\citep{JackAmb}. Several different approaches have
been developed to remove this ambiguity (see for example\,\citep{Brito:2007uc,DelCima:2009ta}
and references therein), and the conclusions seems to be that, in
this particular case, gauge invariance enforces the generated CFJ
term to vanish. In any case, this very first example in the study
of quantum corrections induced by the LV coefficients of the SME already
shows the complexity and the potential for interesting theoretical
and phenomenological discussions related to this matter.

These calculations naturally motivated many other investigations,
where specific LV operators where generated as a radiative corrections.
Restriction to the minimal SME allows for more systematical studies
such as the ones presented in\,\citep{KostPic}, since renormalizability
greatly reduces the number of allowed terms in the effective action.
When the extension to the non-minimal SME is considered, operators
with mass dimension greater than four are included, which are non
renormalizable by power counting, so that the theory can only be understood
as an effective field theory. Still, one can look for specific cases
where non minimal LV operators can induce interesting results in the
quantum theory. It is clear that the case of finite and non ambiguous
LV corrections is particularly interesting, since it could relate
LV couplings from different sectors of the framework, and present
itself as a consistent way to generate LV from some more \textquotedbl fundamental\textquotedbl{}
setup (assuming for instance the integrated field not to be one of
the fields in the Standard Model). When divergent corrections to a
given LV operator are generated, this operator has to be introduced
from the very beginning to act as a counterterm to cancel this divergence,
leaving behind an arbitrary finite constant that has to be fixed by
some physical condition. In this case, as in the \textquotedbl finite
but undetermined\textquotedbl{} scenario described previously, one
cannot claim to have generated a given LV operator without some degree
of ambiguity, thus rendering less clear any phenomenological application
of this result. Even so, this approach have been used in literature
to infer bounds on LV coefficients, either by implicitly assuming
all finite constants to be of order one, or by fixing it using minimal
subtraction\,\citep{Anderson:2004qi,Casana:2013nfx}. 

Interestingly enough, it has been shown that, in certain cases, even
the non-renormalizable couplings can yield finite, well-defined results.
The first known example of such a situation is the generation of a
finite aether term from the magnetic Lorentz-breaking coupling\,\citep{aether}.
This is a motivation for further examining quantum corrections in
the non-minimal extensions of QED. Among the interesting results in
this context one can mention, for example, the generation of the axion
term from the Lorentz-breaking couplings \citep{axion} and the CPT-even
terms proportional to the fourth-rank tensor \citep{Casana:2013nfx}.
It is noteworthy to mention that LV operators can also be generated
by other mechanisms, such as spacetime varying couplings constants,
either at the classical level\,\citep{Kostelecky:2002ca}, or in
the computation of loop corrections\,\citep{Ferrero:2009jb}

In this work, we revisit the question of the perturbative generation
of quantum corrections in various sectors of SME, revisiting the relevant
literature and presenting some new results. One objective is to put
in a more general and systematic perspective the problem of radiative
corrections in the SME. We will fill some gaps in the literature,
in the sense of presenting the results in the standard SME notation,
which greatly facilitates the comparison with experimental results.
In doing so, we will show that new and interesting information can
be obtained. For example, we will argue that higher order corrections
induced from the $b^{\mu}$ coefficient in the birefringent part of
the $k_{F}$ term of the photon sector may provide stronger constraints
on the temporal component of $b^{\mu}$ than the ones currently known.
We therefore provide a \textquotedbl road map\textquotedbl{} for
obtaining more results of this kind while, on the same time, motivating
the adherence to the SME notation in these studies. 

The structure of the paper looks like follows. In section\,\ref{sec:minimalQED},
we look at the minimal interactions between gauge and spinor fields,
and the corresponding operators generated in the gauge sector, extending
this study for the non minimal QED extension in section\,\ref{sec:nonminimalQED}.
In section\,\ref{sec:Spinor-scalar-LV-couplings}, we carry the same
study for scalar-spinor couplings, and in section\,\ref{sec:LV-spinor}
for the contributions with external spinors. Section\,\ref{sec:Gravity}
is devoted to a discussion of Lorentz-breaking quantum corrections
in a curved space-time. Finally, section\,\ref{sec:Conclusions}
contains our conclusions and final remarks.

\section{The Minimal QED Extension\label{sec:minimalQED}}

The most generic Lorentz-breaking extension of QED containing only
terms of renormalizable dimensions, also called the minimal QED extension,
is given by the following Lagrangian\,\citep{coll2}, 
\begin{equation}
{\cal L}=\bar{\psi}(i\Gamma^{\nu}D_{\nu}-M)\psi+{\cal L}_{photon}^{(3,4)}\thinspace,\label{genrenmod}
\end{equation}
where \begin{subequations} 
\begin{align}
\Gamma^{\nu} & =\gamma^{\nu}+c^{\mu\nu}\gamma_{\mu}+d^{\mu\nu}\gamma_{\mu}\gamma_{5}+e^{\nu}+if^{\nu}\gamma_{5}+\frac{1}{2}g^{\lambda\mu\nu}\sigma_{\lambda\mu}\thinspace,\\
M & =m+a_{\mu}\gamma^{\mu}+b_{\mu}\gamma^{\mu}\gamma_{5}+\frac{1}{2}H^{\mu\nu}\sigma_{\mu\nu}\thinspace,
\end{align}
\end{subequations}$D_{\mu}=\partial_{\mu}-ieA_{\mu}$ is the usual
$U\left(1\right)$ covariant derivative, ${\cal L}_{photon}^{(3,4)}$
is the restriction of the Lagrangian in Eq.\,\eqref{eq:Lph} to minimal
(dimension three and four) operators, and $a^{\mu}$, $b^{\mu}$,
$c^{\mu\nu}$, $d^{\mu\nu}$, $e^{\mu}$, $f^{\mu}$, $g^{\lambda\mu\nu}$,
$H^{\mu\nu}$ are constant (pseudo)tensors, collectively known as
the LV coefficients of the QED sector of the minimal SME, together
with the $\kappa_{F}^{(4)}$ and $k_{AF}^{(3)}$ defined by Eq.\,\eqref{eq:genKAF}.

While generality motivates equation\,\eqref{genrenmod}, incorporating
into the QED Lagrangian all possible terms respecting gauge invariance,
power counting renormalizability and observer Lorentz invariance,
one should be aware that some of the LV couplings present in\,\eqref{genrenmod}
are actually not relevant to physics in most cases. A first example
is the vector $a_{\mu}$, which can be eliminated in a single fermion
theory by a field redefinition $\psi\rightarrow e^{-ia\cdot x}\chi$.
While in a theory with different fermion species and different $a$'s
one might find some non trivial effects arising from this coefficient,
it is usually disregarded; the situation changes when gravity is taken
into account, however\,\citep{KosGra}. The antisymmetric part of
the $c^{\mu\nu}$ coefficient can be removed by a redefinition of
the gamma matrices, so $c^{\mu\nu}$ is usually taken to be symmetric;
also, the antisymmetric part of $d^{\mu\nu}$, the trace and totally
antisymmetric parts of the $g^{\lambda\mu\nu}$ terms are not expected
to generate independent physical effects\,\citep{coll2}. The $f^{\mu}$
coefficient can be shown to generate effects that can be exactly mimicked
by the symmetric part of $c^{\mu\nu}$\,\citep{Altschul:2006ts},
so it is also usually disregarded. A detailed discussion of the removal
of spurious LV operators via field redefinitions can be found in\,\citep{Colladay:2002eh}
(see also\,\citep{Lehnert:2006id} for a related discussion on the
$b^{\mu}$ coefficient, and\,\citep{Ferrari:2016szq} for a discussion
involving singular spinor fields and torsion).

The photon sector of the SME being so well understood theoretically,
together with the strong experimental constraints that can be obtained
from it, makes particularly interesting the study of the structure
of quantum corrections that can be induced from different LV couplings
in this specific sector of the SME. For the remainder of this section,
restrict ourselves to the original couplings being present in the
minimal QED extension, as described in\,\eqref{genrenmod} (notice,
however, that in several instances, the generated LV operators are
themselves non minimal), and discuss the most interesting cases of
radiative corrections. Whenever relevant, we will rewrite the original
results obtained in the literature in the standard SME notation and
comment on possible phenomenological bounds that might be inferred
from these results, an analysis which is mostly missing on many of
those works. 

Maybe the most studied instance of the mechanism we are interested
in, as we already mentioned in the introduction, is the one involving
the axial vector $b_{\mu}$, mainly due to the fact the relevant calculation
is puzzled by a technical ambiguity. Specifically, we are interested
in the corrections induced in the photon sector from the LV minimal
operator 
\begin{equation}
V_{b}=b_{\mu}\bar{\psi}\gamma^{\mu}\gamma_{5}\psi\thinspace,\label{eq:Vb}
\end{equation}
after integration of the fermion loop. This LV insertion in a fermion
loop contributing to the two-point photon vertex function generates
the CFJ term\,\citep{coll2,JK}, 
\begin{equation}
{\cal L}_{eff}\supset C_{0}e^{2}\epsilon^{\mu\nu\lambda\rho}b_{\mu}A_{\nu}\partial_{\lambda}A_{\rho}\thinspace,\label{cfj}
\end{equation}
$e$ being the electric charge. This amounts, in the SME notation,
to the generation of a minimal $k_{AF}$ term with $k_{AF}\sim b$.
In this result, however, $C_{0}$ is a finite and ambiguous constant.
The ambiguity is no surprise since it comes from a triangular fermion
loop with one insertion of the LV two-fermion vertex $b_{\mu}\gamma^{\mu}\gamma_{5}$,
whose corresponding amplitude is therefore quite similar to the well
known anomalous triangle diagram in the Standard Model, where one
of the external photon lines is taken at zero external momenta: $A\left(p\right)_{\mu}\gamma^{\mu}\gamma_{5}\rightarrow A\left(0\right)_{\mu}\gamma^{\mu}\gamma_{5}$.
It has also been shown that in the non-Abelian case, for $N>1$ spinor
fields, the $N$-fields generalization of the $V_{b}$ operator generates
the non-Abelian extension of the CFJ term\,\citep{nab}, 
\begin{equation}
{\cal L}_{eff}\supset C_{0}\epsilon^{\mu\nu\lambda\rho}b_{\mu}{\rm tr}\left(g^{2}A_{\nu}\partial_{\lambda}A_{\rho}+\frac{2g^{3}}{3}A_{\nu}A_{\lambda}A_{\rho}\right)\thinspace,
\end{equation}
$C_{0}$ being the same ambiguous constant. Several works in the literature
claim that gauge invariance actually enforces $C_{0}=0$\,\citep{ColGlas,Bonneau1},
so this mechanism can hardly be argued to generate in a consistent
way the minimal $k_{AF}$ term in the photon sector of the SME.

One might also study higher order corrections derived from the $V_{b}$
insertion, both in the sense of an expansion in derivatives of the
electromagnetic field, as well as higher orders in the LV coefficient.
In the first sense, keeping only one $V_{b}$ insertion, but calculating
higher derivative contributions, the result yields the higher-derivative
CFJ-like term 
\begin{equation}
{\cal L}_{eff}\supset C_{1}\frac{e^{2}}{m^{2}}\epsilon^{\beta\mu\nu\rho}b_{\beta}A_{\mu}\Box F_{\nu\rho}\thinspace,\label{HDCFJ}
\end{equation}
with $C_{1}=1/24\pi^{2}$ being a finite, well defined constant\,\citep{TMariz2}.
In the SME notation, this amounts to the generation of a dimension
five coefficient 
\begin{equation}
(\hat{k}_{AF}^{\left(5\right)})_{\kappa}^{\hphantom{\kappa}\alpha\beta}=C_{1}\frac{e^{2}}{2m^{2}}b_{\kappa}\eta^{\alpha\beta}\thinspace.\label{eq:HDCFJkAF}
\end{equation}
The interesting aspect of this calculation is that the result is ambiguity-free,
so one could hope to infer experimental bounds on $b$ from the constraints
on the photon sector coefficient $\hat{k}_{AF}^{\left(5\right)}$.
However, at the moment, experimental constraints on dimension five
photon coefficients are obtained only by the study of free propagation
of photons\,\citep{datatables}, which means that leading LV effects
may be obtained by imposing the usual dispersion relation $\eta_{\mu\nu}p^{\mu}p^{\nu}=0$
in the general expressions for $\hat{k}_{AF}$ given in Eq.\,\eqref{eq:genKAF}.
This, together with Eq.\,\eqref{eq:HDCFJkAF}, means that the LV
operator in Eq.\,\eqref{HDCFJ} does not contribute to wave propagation
(the same result can be seen simply by the fact that $\Box F^{\mu\nu}=0$
for free photon propagation, at the leading order). As a conclusion,
in this example, even if the generated LV operator is finite and free
of ambiguities, its particular form is such that no experimental constraints
can be inferred from this result at the present.

Despite $V_{b}$ involving an assumedly very small LV coefficient,
it might be still interesting to look for corrections of higher order
in $b$, since these may provide unique effects, which might not be
obscured from lower order corrections. In particular, at second order,
$V_{b}$ can contribute to the CPT-even, minimal, $k_{F}$ coefficient.
Considering two $V_{b}$ insertions, one indeed obtains the aether
term, 
\begin{equation}
{\cal L}_{eff}\supset-C_{2}\frac{e^{2}}{m^{2}}b^{\mu}b_{\lambda}F_{\mu\nu}F^{\lambda\nu}\thinspace,\label{aether}
\end{equation}
with $C_{2}=1/6\pi^{2}$. This result is finite by power counting
and therefore ambiguity-free\,\citep{Bonneau,aether2}, and corresponds
to the SME minimal coefficient 
\begin{equation}
(k_{F}^{(4)})^{\mu\nu\alpha\beta}=-C_{2}\frac{e^{2}}{m^{2}}\left(b^{\mu}b^{\alpha}\eta^{\nu\beta}-b^{\nu}b^{\alpha}\eta^{\mu\beta}-b^{\mu}b^{\beta}\eta^{\nu\alpha}+b^{\nu}b^{\beta}\eta^{\mu\alpha}\right)\thinspace.\label{eq:aetherkF}
\end{equation}
Since this is a well defined quantum correction, we may question whether
this result can induce competitive constraints on the $b_{\mu}$ coefficients,
given that very strong constraints exist on $k_{F}$ due to birefringence
effects. One simple parametrization for the components of $k_{F}$
relevant for birefringence is in terms of the ten $k^{a}$ coefficients
defined in\,\citep{Kostelecky:2001mb}, which are constrained at
the order $10^{-37}$ from the study of birefringence in gamma ray
bursts\,\citep{Kostelecky:2006ta}. Therefore, from Eq.\,\eqref{eq:aetherkF}
we could expect components of $b$ to be limited by $b^{2}<6\pi^{2}m^{2}/e^{2}\times10^{-37}$,
amounting to $\left|b_{p}\right|<3\times10^{-15}\thinspace\text{GeV}$
for protons and $\left|b_{e}\right|<1.5\times10^{-20}\thinspace\text{GeV}$
for electrons, for example: these are not better than the limits already
known for the spatial components of $b_{\mu}$, but are potentially
better than the ones for the temporal components of $b_{\mu}$, which
are $\left|(b_{T})_{p}\right|<7\times10^{-8}\thinspace\text{GeV}$
for protons and $\left|(b_{T})_{e}\right|<10^{-15}\thinspace\text{GeV}$
for electrons\,\citep{datatables}, where the $T$ subscript is the
standard notation for the temporal component of a vector in the Sun
centered frame. To verify this, we assume for the moment that $b^{\mu}=\left(b,\vec{0}\right)$
in the Sun centered frame, which is adopted as the standard frame
of reference for quoting experimental constraints on the SME coefficients,
and we take into account that quantum corrections will induce a corresponding
$(k_{F})^{\mu\nu\alpha\beta}$ of the form given in Eq.\,\eqref{eq:aetherkF}.
It is easy to verify that we generate this way non-vanishing birefringent
coefficients $k^{a}=C_{2}e^{2}b^{2}/m^{2}$ for $a=3,4$, which are
subjected to the aforementioned $10^{-37}$ constraints. Alternatively,
one may obtain the matrices $\kappa_{DE}$, $\kappa_{DB}$ and $\kappa_{HB}$
as defined in\,\citep{KosteleckyMewes2002} as $\kappa_{DE}=-\left(2C_{2}e^{2}b^{2}/m^{2}\right){\bf 1}$
and $\kappa_{DB}=\kappa_{HB}={\bf 0}$, and from this, $\tilde{\kappa}_{e\pm}=\mp\left(C_{2}e^{2}b^{2}/m^{2}\right){\bf 1}$,
$\tilde{\kappa}_{o\pm}=0$ and $\tilde{\kappa}_{{\rm tr}}=-2C_{2}e^{2}b^{2}/m^{2}$,
and the birefringent matrix $\tilde{\kappa}_{e+}$ is to be subjected
to the $10^{-37}$ limit. From this argument, we conclude that the
consistency of the quantum corrections given by Eq.\,\eqref{aether}
imply in the following constraints for the temporal component of the
$b_{\mu}$ pseudo-vector, written in general as 
\begin{equation}
\left|b_{T}\right|<\pi m/e\sqrt{6\times10^{-37}}\thinspace.
\end{equation}
This implies in new stronger experimental constraints on $b_{T}$
coefficients, for example
\begin{equation}
\left|(b_{T})_{p}\right|<3\times10^{-15}\thinspace\text{GeV}
\end{equation}
for protons and 
\begin{equation}
\left|(b_{T})_{e}\right|<1.5\times10^{-20}\thinspace\text{GeV}
\end{equation}
 for electrons. It is interesting to note that we are able to obtain
these new constraints, despite the induced operator being of the second
order in $b_{\mu}$, and this happens because one of the components
of $b_{\mu}$ (the temporal one) have not been so tightly constrained
so far. 

Higher orders corrections in $V_{b}$ are not expected to lead to
competitive bounds, yet they have been studied for theoretical reasons.
Considering three $V_{b}$ insertions, one obtains a linear combination
of the higher-derivative CFJ-like term (\ref{HDCFJ}) (with a coefficient
proportional to $b^{2}$) and the Myers-Pospelov term\,\citep{MP}
\begin{equation}
{\cal L}_{eff}\supset\frac{e^{3}}{m^{4}}C_{3}b^{\alpha}F_{\alpha\mu}(b\cdot\partial)\epsilon^{\beta\mu\nu\rho}b_{\beta}F_{\nu\rho}\thinspace,\label{MP}
\end{equation}
where $C_{3}$ is also an ambiguity-free constant\,\citep{MNP}.
This model can be obtained from the general SME formalism by a specific
choice of $\hat{k}_{AF}^{\left(5\right)}$, corresponding to a particular
isotropic limit of Lorentz violation, leading to modified dispersion
relations for photons (which were the original motivation for the
introduction of the model in\,\citep{MP}), see section IV-F in\,\citep{KostelMewes-EMHD}
for more details. It is interesting to note that the higher-derivative
CFJ-like term generated in this case does not appear for a light-like
vector $b_{\mu}$. For more insertions, it is natural to expect the
appearance of terms including fourth and higher orders in derivatives,
meaning contributions to $k_{F}^{(d)}$ for $d\geq6$. Up to now,
apart from the general discussion in\,\citep{KostelMewes-EMHD},
further consequences of these dimension six terms have only been studied
at the tree level\,\citep{Casana:2018rhg}.

After this discussion of the quantum consequences of the $V_{b}$
operator, we will consider the other relevant minimal couplings. We
start with the operator proportional to the tensor $c_{\mu\nu}$,
\begin{equation}
V_{c}=ic^{\mu\nu}\bar{\psi}\gamma_{\mu}(\partial_{\nu}-ieA_{\nu})\psi\thinspace.
\end{equation}
Without loss of generality, $c^{\mu\nu}$ is taken to be symmetric,
as discussed before. A minimal scenario including only the $c^{\mu\nu}$
coefficient was discussed in\,\citep{Schaposnik}, where its presence
was shown to be equivalent to the redefinition of the Dirac matrices
through 
\begin{equation}
\gamma^{\mu}\to\gamma^{\mu}+c^{\mu\nu}\gamma_{\nu}\thinspace,
\end{equation}
with the subsequent arising of the deformed metric 
\begin{equation}
M^{\mu\nu}=(\delta_{\thinspace\alpha}^{\mu}+c_{\thinspace\alpha}^{\mu})(\delta_{\thinspace\beta}^{\nu}+c_{\thinspace\beta}^{\nu})\eta^{\alpha\beta}\thinspace.
\end{equation}
As a result, one can arrive at quantum corrections involving contractions
of the gauge field $A_{\mu}$ and the stress tensor $F_{\mu\nu}$
with the constant tensors $\delta_{\alpha}^{\mu}+c_{\alpha}^{\mu}$.
The Adler-Bell-Jackiw anomaly was the main interest of the reference\,\citep{Schaposnik},
and it was shown that the presence of the $c^{\mu\nu}$ coefficient
as the only LV in the model does not modify the usual picture of the
anomaly and index theorem. As for the radiative generation of corrections
in the photon sector, calculations have been performed only considering
a particular form of the $c_{\mu\nu}$ coefficient, parametrized by
a constant vector $u_{\mu}$, 
\begin{equation}
c_{\mu\nu}=u_{\mu}u_{\nu}-\frac{\zeta}{4}\eta_{\mu\nu}u^{2}\thinspace,\label{cmn}
\end{equation}
so that, at $\zeta=0$ we have the simplest form $c_{\mu\nu}=u_{\mu}u_{\nu}$,
and at $\zeta=1$ the $c_{\mu\nu}$ is traceless. Respecting CPT invariance,
quantum corrections involving the $V_{c}$ insertion will contribute
to the CPT-even photon coefficient $k_{F}$, starting at the first
order, and the explicit form of this contribution involving up to
three $c_{\mu\nu}$ insertions has been calculated in\,\citep{Maluf}.
The most interesting case to quote is the first order in the traceless
$c_{\mu\nu}$, where the leading (divergent) corrections generate
the simple term
\begin{equation}
{\cal L}_{eff}\supset\frac{e^{2}}{2\pi^{2}\epsilon}c_{\mu\rho}\eta_{\nu\sigma}F^{\mu\nu}F^{\sigma\rho}\thinspace,
\end{equation}
corresponding to 
\begin{equation}
(k_{F})_{\mu\nu\rho\sigma}\sim c_{\mu\rho}\eta_{\nu\sigma}-c_{\nu\rho}\eta_{\mu\sigma}-c_{\mu\sigma}\eta_{\nu\rho}+c_{\nu\sigma}\eta_{\mu\rho}\thinspace.
\end{equation}
For $\zeta=0$, one obtains in particular the aether-like photon coefficient
in Eq.\, (\ref{eq:aetherkF}), together with a rescaled Maxwell term
proportional to $u^{2}F^{\mu\nu}F_{\mu\nu}$. However, unlike in the
results generated from the CPT-odd couplings\,\citep{Bonneau,aether2},
in the case of the $c_{\mu\nu}$ insertions the aether term logarithmically
diverges. 

The term proportional to the coefficient $d_{\mu\nu}$ is 
\begin{equation}
V_{d}=id^{\mu\nu}\bar{\psi}\gamma_{\mu}\gamma_{5}(\partial_{\nu}-ieA_{\nu})\psi\thinspace.
\end{equation}
Early works concerning this term include\,\citep{KostPic}, where
the study of one-loop renormalizability of the extended QED involved
the calculation of divergent corrections induced by all the minimal
coefficients in\,\eqref{genrenmod}, and also\,\citep{Schaposnik},
where it was shown that compatibility with chiral symmetry allows
for a $d_{\mu\nu}$ coefficient which is not independent, but actually
defined in terms of the $c_{\mu\nu}$ coefficient by $d_{\thinspace\nu}^{\mu}=Q(\delta_{\thinspace\nu}^{\mu}+c_{\thinspace\nu}^{\mu})$,
leading to the particular QED extension involving only the $c^{\mu\nu}$
independent coefficient that was already mentioned in the previous
paragraph. 

From a more general perspective, the $d_{\mu\nu}$ is CPT even and
therefore can only contribute to the CPT even photon coefficient $k_{F}$,
however, since $d_{\mu\nu}$ is a pseudotensor, the possible LV contributions
generated in the photon sector will involve even orders in $d_{\mu\nu}$,
starting from a minimal term of the general form $d^{\mu\alpha}d^{\nu\beta}F_{\mu\nu}F_{\alpha\beta}$,
corresponding to the generation of a $k_{F}$ term with 
\begin{equation}
(k_{F})^{\mu\nu\alpha\beta}\sim d^{\mu\alpha}d^{\nu\beta}-d^{\nu\alpha}d^{\mu\beta}\thinspace.
\end{equation}
A first-order term, whose only possible structure respecting observer
Lorentz invariance would be like $\epsilon_{\mu\nu\alpha\rho}F^{\mu\nu}d_{\lambda}^{\alpha}F^{\lambda\rho}$,
would correspond to $(k_{F})^{\mu\nu\alpha\beta}\sim\epsilon^{\mu\nu\alpha\rho}d_{\alpha}^{\thinspace\lambda}$,
which does not possess the necessary symmetry properties except if
$d_{\thinspace\nu}^{\alpha}\sim\delta_{\thinspace\nu}^{\alpha}$,
which is evidently trivial since there is no breaking of Lorentz symmetry
in this case; moreover, the absence of first order in $d_{\mu\nu}$
corrections has been verified through direct calculations\,\citep{KostPic}.
The second order contribution is divergent: actually, its pole part
has been shown in\,\citep{Scarp3} to possess the same structure
as the second order in $c_{\mu\nu}$ corrections found in\,\citep{Maluf}.
Notice however that explicit results have only been obtained so far
for $c_{\mu\nu}$ only for the particular form in Eq.\,\eqref{cmn},
while $d_{\mu\nu}$ being a pseudotensor, it cannot be cast in the
same form, and therefore, no explicit results for the calculation
of the $d_{\mu\nu}$ corrections are available. 

Despite $V_{c}$ contributing to $k_{F}$ already at the first order,
and $V_{d}$ at the second order, a phenomenological analysis of these
results is obscured by the fact that these corrections are divergent,
and that a particular choice of these tensors have been used in the
literature to obtain explicit results. 

The term proportional to $e^{\mu}$ can contribute in the quantum
corrections starting in the second order: being a vector instead of
a pseudo-vector, it cannot be used at first order to construct the
$k_{AF}$ term, and being a CPT-odd term, it can only contribute to
$k_{F}$ at second order. The same applies to $f^{\mu}$, as can be
checked through straightforward calculations\,\citep{Scarp3}. It
turns out that both corrections have exactly the same form, amounting
to divergent aether-like corrections like the ones in\,(\ref{aether}).
Since the calculations in\,\citep{Scarp3} where performed with an
implicit regularization method, we can quote the explicit result for
the generated corrections to $k_{F}$ as
\begin{equation}
(k_{F})^{\mu\nu\rho\sigma}=\frac{e^{2}}{12}I_{log}\left(m^{2}\right)\left(\eta^{\rho\mu}e^{\nu}e^{\sigma}-\eta^{\rho\nu}e^{\mu}e^{\sigma}-\eta^{\sigma\mu}e^{\nu}e^{\rho}+\eta^{\sigma\nu}e^{\mu}e^{\rho}\right)\thinspace,
\end{equation}
with the corresponding expression for $f^{\mu}$ being obtained by
simply substituting $e^{\mu}$ by $f^{\mu}$, and where $I_{log}\left(m^{2}\right)$
is a logarithmically divergent expression that may be calculated in
different regularization schemes. In dimensional regularization, for
example, one has
\begin{equation}
I_{log}\left(m^{2}\right)=\frac{i}{16\pi^{2}}\Gamma\left(\frac{\epsilon}{2}\right)\left(\frac{4\pi m^{2}}{\mu^{2}}\right)^{\epsilon/2}\thinspace.
\end{equation}

The term proportional to $g^{\mu\nu\lambda}$ will yield the finite
and well defined higher-derivative contribution\,\citep{MarizHD}
\begin{equation}
{\cal L}_{eff}\supset\frac{e^{2}}{24m\pi^{2}}{\cal G}^{\mu\nu\rho\alpha\beta}A_{\alpha}\partial_{\rho}\partial_{\alpha}\partial_{\beta}A_{\nu}\thinspace,\label{eq:Lg}
\end{equation}
$e$ being the charge and $m$ the mass of the integrated fermion,
and 
\begin{equation}
{\cal G}^{\mu\nu\rho\alpha\beta}=g^{\mu\nu\alpha}\eta^{\rho\beta}+g^{\mu\nu\beta}\eta^{\rho\alpha}-g^{\mu\rho\alpha}\eta^{\nu\beta}-g^{\mu\rho\beta}\eta^{\nu\alpha}-g^{\rho\nu\alpha}\eta^{\mu\beta}-g^{\rho\nu\beta}\eta^{\mu\alpha}\thinspace.
\end{equation}
In the SME notation, this amounts to a contribution to $\hat{k}_{AF}^{(5)}$
of the form 
\begin{equation}
(\hat{k}_{AF}^{\left(5\right)})_{\kappa}^{\hphantom{\kappa}\alpha\beta}=\frac{e^{2}}{24\times3!\thinspace m\pi^{2}}\epsilon_{\kappa\lambda\mu\nu}{\cal G}^{\lambda\mu\nu\alpha\beta}\thinspace.
\end{equation}
Despite being finite and well defined, this correction actually does
not contribute to photon propagation in leading order. This can be
seen by noticing that either the leading order dispersion relation\footnote{This dispersion relation can be obtained from the general results
presented in\,\citep{KostelMewes-EMHD}, or by the appropriate limit
of the dispersion relation presented in\,\citep{MarizHD}, which
also includes a subleading term involving a $p^{2}\left(k_{AF}\right)^{2}$
factor.}
\begin{equation}
\left(p^{2}\right)^{2}-4\left(p^{\mu}\left(k_{AF}\right)_{\mu}\right)^{2}\approx0
\end{equation}
or the relevant Stokes parameter\footnote{See section IIIC of\,\citep{KostelMewes-EMHD}.}
\begin{equation}
\varsigma^{3}=-p^{\mu}\left(k_{AF}\right)_{\mu}/\omega^{2}
\end{equation}
are modified by the combination $p^{\mu}\left(k_{AF}\right)_{\mu}$,
which can be shown to vanish. Indeed, from the antisymmetry of $g^{\mu\nu\lambda}$
in the first two indices, it can be shown that
\begin{equation}
(k_{AF})_{\mu}=(\hat{k}_{AF}^{\left(5\right)})_{\mu}^{\hphantom{\kappa}\alpha\beta}p_{\alpha}p_{\beta}=6\epsilon_{\mu\nu\rho\sigma}g^{\nu\rho\alpha}p_{\alpha}p^{\sigma}\thinspace,
\end{equation}
and therefore $p^{\mu}(k_{AF})_{\mu}=0$ as a result of the contraction
of the epsilon with two momenta. Current limits on dimension five
photon coefficients all derive from astrophysical observations of
photon propagation and, therefore, cannot be used to impose limits
on $g^{\mu\nu\lambda}$ based on the induced term\,\eqref{eq:Lg}.

Finally, we note that for the particular case of completely antisymmetric
$g_{\mu\nu\lambda}=\epsilon_{\mu\nu\lambda\rho}h^{\rho}$, Eq.\,\eqref{eq:Lg}
yields the finite higher-derivative CFJ-like result\,(\ref{HDCFJ}),
with an appropriate multiplying factor. The finite temperature behavior
of this term is discussed in\,\citep{TMariz1}. In the second order
in $g_{\mu\nu\lambda}$, for the same case of a completely antisymmetric
$g_{\mu\nu\lambda}$, one arrives at the logarithmically divergent
aether-like result\,(\ref{aether}), with $b_{\mu}$ replaced by
$h_{\mu}$. 

To close the discussion of the minimal part, it remains to discuss
the impacts of the $H_{\mu\nu}$ term. One can naturally make the
conclusion that the lower possible contribution involving this insertion
should be at least of the second order (the first-order contribution
evidently vanishes by symmetry reasons), and it must be superficially
finite by dimensional arguments. It is natural to expect expression
of the form $H^{\mu\nu}H^{\alpha\beta}F_{\mu\alpha}F_{\nu\beta}$.
However, explicit calculation shows that this term identically vanishes
at the one-loop order\,\citep{Scarp3}.

\section{Non-minimal extensions of QED\label{sec:nonminimalQED}}

It is very natural to study quantum corrections in the minimal SME,
which is proven to be a renormalizable model. In a more general perspective,
however, the SME is an effective field theory which naturally includes
non-minimal operators of mass dimension greater than four, naturally
appearing in various phenomenological models (see f.e. \citep{Antoniadis:2008es}),
and in this section we want to discuss the effects of these in the
quantum corrections.

From the formal viewpoint, the presence of such couplings is a natural
consequence of the fact that the SME is an effective field theory
arising in the low-energy limit of some fundamental theory at a very
high energy scale $\Lambda$, therefore it depends on this characteristic
energy scale, with non-minimal vertices being proportional to negative
powers of $\Lambda$\,\citep{Georgi}. While the restriction to dimension
three and four operators, corresponding to the minimal SME, leads
to a consistent quantum field theory by itself (which, being renormalizable,
is actually independent of the scale $\Lambda$), the general picture
is certainly less clear for the non minimal SME, since higher-dimension
kinetic operators, due to the presence of higher derivatives, typically
yield ghost excitations, while higher-derivative interactions are
essentially non-renormalizable. So, it is not expected that consistent
quantum corrections can be calculated in general, however specific
terms can be shown to provide interesting results, and indeed several
examples have been reported in the literature so far. Up to now, most
of these studies focused on the leading, dimension-five operators,
with the dimension six case being discussed recently in\,\citep{Casana:2018rhg}
for the gauge sector, and in\,\citep{CMR1} for the spinor sector.

The first non-minimal coupling whose quantum impacts were studied
at the perturbative level is the dimension five magnetic one\,\citep{aether},
involving a single LV vector $u_{\beta}$, 
\begin{equation}
{\cal V}_{1}=gu_{\beta}\thinspace\bar{\psi}\gamma_{\alpha}\psi\thinspace\epsilon^{\alpha\beta\gamma\delta}F_{\gamma\delta}\thinspace,\label{v1n}
\end{equation}
or, in the SME notation\,\citep{Ding:2016lwt}, 
\begin{equation}
{\cal V}_{1}=-\frac{1}{2}(a_{F}^{(5)})^{\alpha\beta\gamma}\bar{\psi}\gamma_{\alpha}\psi F_{\beta\gamma}\thinspace,\label{eq:aF5}
\end{equation}
where $(a_{F}^{(5)})^{\alpha\beta\gamma}=2g\thinspace\epsilon^{\rho\alpha\beta\gamma}u_{\rho}$.
There are at the moment no experimental constraints reported on these
non-minimal coefficients\,\citep{datatables}. It is interesting
to notice that in the works mentioned by us, $u^{\beta}$ is assumed
to be a mass dimension one LV coefficient, just as the minimal extended
QED coefficient $b^{\beta}$ (actually, in many instances, the coupling
$V_{b}$ is also considered in the calculation, and it is assumed
that $u^{\beta}=b^{\beta}$), so that $g$ has mass dimension $-2$.
One remarkable fact related to this vertex is that the contribution
to the two-point function of the gauge field generated by two such
vertices, although quadratically divergent by power counting, unexpectedly
yields a finite aether-like result 
\begin{equation}
{\cal L}_{eff}\supset C_{4}m^{2}g^{2}\thinspace u^{\mu}u_{\lambda}F_{\mu\nu}F^{\lambda\nu}\thinspace,\label{aethres}
\end{equation}
$C_{4}$ being an ambiguous dimensionless finite constant studied
in detail in\,\citep{aether2}. When this mechanism is considered
at finite temperature, the situation becomes more involved, for example,
an aether-like term involving only spatial components like $u_{i}u^{k}F^{ij}F_{kj}$
becomes possible\,\citep{aetherT}. The non-Abelian generalization
of this calculation is possible as well. The ambiguity in the calculation
of the constant $C_{4}$ in principle precludes a confident phenomenological
analysis, with the objective of transferring the bounds on $(k_{F})^{\mu\nu\alpha\beta}=-4C_{4}m^{2}g^{2}\left(u^{\mu}u^{\lambda}\eta^{\nu\rho}-u^{\nu}u^{\lambda}\eta^{\mu\rho}-u^{\mu}u^{\rho}\eta^{\lambda\rho}+u^{\nu}u^{\rho}\eta^{\mu\lambda}\right)$,
which corresponds in the SME notation to Eq.\,\eqref{aethres}, to
a bound in $a_{F}^{(5)}$, which would be a very interesting result.

Other corrections arising from the presence of the ${\cal V}_{1}$
vertex are possible. If one consider the contribution involving one
${\cal V}_{1}$ and one usual vertex $-e\bar{\psi}\slashed{A}\psi$,
for example, the CFJ term proportional to the same ambiguous constant
$C_{4}$ would be obtained. Calculating the higher-derivative contributions
to the two-point function generated by Feynman diagrams involving
either two non-minimal ${\cal V}_{1}$ vertices or one ${\cal V}_{1}$
and one usual QED vertex, with one or two minimal $V_{b}$ insertions,
the result will be superficially finite, being a linear combination
of the Myers-Pospelov term\,(\ref{MP}) and the higher-derivative
CFJ term\,(\ref{HDCFJ}), just as it occurs for the case when both
vertices are minimal\,\citep{MNP}. One of the contributions to each
of these terms will be ambiguous. The complete result for the linear
combination of these terms, generated by the presence of both interactions,
contains\,(\ref{MP}) as one of the contributions, and looks like
\begin{align}
{\cal L}_{eff}\supset & \left(2g^{2}C_{1}+\frac{eg}{6\pi^{2}m^{2}}+\frac{4e^{2}}{45\pi^{2}m^{4}}\right)u^{\alpha}F_{\alpha\mu}(b\cdot\partial)u_{\beta}\epsilon^{\beta\mu\nu\lambda}F_{\nu\lambda}\nonumber \\
 & +\left(2g^{2}C_{1}+\frac{eg}{6\pi^{2}m^{2}}+\frac{e^{2}}{9\pi^{2}m^{4}}\right)u^{2}u_{\beta}\epsilon^{\beta\mu\nu\lambda}A_{\mu}\Box F_{\nu\lambda}\thinspace,
\end{align}
where $C_{1}$ is the same finite and ambiguous constant involved
in the generation of the CFJ term\,(\ref{cfj}), as described in
the previous section. Again, the remarkable property is the finiteness
of this result, despite the initial power counting of the Feynman
diagrams involved. 

Another non-minimal, CPT odd vertex have been discussed in the literature
in connection with axion physics, to wit, 
\begin{equation}
{\cal V}_{2}=v^{\beta}\thinspace\bar{\psi}\gamma^{\alpha}\psi\thinspace F_{\alpha\beta}\thinspace,\label{v2n}
\end{equation}
which is also of the same form as Eq.\,\eqref{eq:aF5}, but with
\begin{equation}
(a_{F}^{(5)})^{\alpha\beta\gamma}=-\left(v^{\beta}\eta^{\gamma\alpha}-v^{\alpha}\eta^{\gamma\beta}\right)\thinspace.
\end{equation}
Notice that, here, $v^{\beta}$ has dimension of inverse of mass.
It has been shown in\,\citep{axion} that a triangle graph similar
to that one studied in\,\citep{JK}, but with one external field
$eA_{\alpha}$ replaced by the ${\cal V}_{2}$ vertex and the insertion
$\slashed{b}\gamma_{5}$ replaced by $\vartheta\slashed{a}\gamma_{5}$,
with $\vartheta=\vartheta(x)$ being the axion field, will generate
in the effective action the usual coupling between the photon and
an axion-like-particle, i.e., 
\begin{equation}
{\cal L}_{eff}\supset C_{1}e\epsilon^{\mu\nu\alpha\beta}b_{\alpha}u^{\rho}\vartheta F_{\mu\nu}F_{\rho\beta}=2C_{1}eg\left(u\cdot b\right)\vartheta\left(\vec{E}\cdot\vec{B}\right)\thinspace,\label{axion}
\end{equation}
where $C_{1}$ is the same ambiguous constant defined in (\ref{cfj}).
In obtaining this result, it was assumed that the integrated fermion
$\psi$ is very massive, so that it makes sense to extract from the
relevant integrals only the dominant results when its mass is very
large compared to any other scale (so it is sufficient to keep the
first term in a derivative expansion for $\vartheta(x)$). Interestingly
enough, this LV mechanism yields an isotropic correction that exactly
mimics the standard axion-photon coupling, which is relevant for many
experimental searches for axion-like-particles\,\citep{axisearch}.
Despite being finite, this calculation suffers from the same sort
of ambiguities present in the generation of the CFJ term. Notice,
however, that the arguments used to argue that $C_{1}$ should vanish,
involving the gauge symmetry of a correction to the photon propagator,
do not necessarily apply in this case, which concerns an interaction
term involving photons and a light pseudoscalar. The possibility of
a phenomenological relevance of the generated term\,\eqref{axion}
was hinted in\,\citep{axion}, however, a proper examination of the
ambiguity in this correction is still missing. Finally, as a comment,
we note that if we replace the magnetic coupling in\,\eqref{v1n}
by the one in\,(\ref{v2n}) within the study of the aether term carried
out in the paper\,\citep{aether}, in the four-dimensional case,
we also will obtain the aether term with the same ambiguous multiplier
$C_{4}$ defined in (\ref{aethres}).

The interaction vertex in\,\eqref{v2n} was further studied in a
series of articles devoted to its quantum effects in the photon sector.
The groundwork was developed in\,\citep{axion2}, considering the
functional determinant 
\begin{equation}
S_{eff}\supset i{\rm Tr}\ln\left(i\slashed{\partial}-e\slashed{A}-m-\gamma^{\alpha}F_{\alpha\beta}v^{\beta}\right)\thinspace,
\end{equation}
which was calculated using the zeta function method, and the result
was expressed in a power series of the electromagnetic field strength.
Wave propagation was studied with the dominant LV corrections that
are generated in the photon sectors, as well as the leading non-linear
corrections, i.e., 
\begin{alignat}{1}
S_{eff}\supset\int d^{4}x\thinspace & \left({\cal L}_{F^{4}}+\frac{g}{12\pi^{2}}\ln\left(\frac{M^{2}}{\mu^{2}}\right)v_{\alpha}F_{\mu\nu}\partial^{\mu}F^{\nu\alpha}\right)\ ,
\end{alignat}
where ${\cal L}_{F^{4}}$ stands for the usual Lorentz invariant Euler-Heisenberg
Lagrangian, with the surprising result that the LV background decouples
from wave propagation in vacuum, in this approximation. It is interesting
to notice that this calculation does not suffer from any ambiguities
of the sort involved in the generation of the CFJ term, yet the leading
LV correction present in the last equation is divergent, thus needing
a renormalization, as indicated by the presence of the renormalization
scale $\mu$. Also, from these results one could extract additional
LV non-linear terms for the field strength, which is a topic still
quite unexplored in the literature.

Now we focus our attention to non-minimal CPT-even couplings. The
first calculation of quantum corrections involving one of these was
presented in\,\citep{Casana:2013nfx}, involving the dimension five
vertex 
\begin{equation}
{\cal V}_{3}=\frac{1}{2}\kappa^{\mu\nu\lambda\rho}\bar{\psi}\sigma_{\mu\nu}\psi F_{\lambda\rho}=-\frac{1}{4}H_{F}^{(5)\mu\nu\alpha\beta}\thinspace\bar{\psi}\sigma_{\mu\nu}\psi F_{\alpha\beta}\thinspace,\label{casana}
\end{equation}
where $H_{F}^{(5)\mu\nu\alpha\beta}=-\frac{1}{2}\kappa^{\mu\nu\lambda\rho}$
in this case, $\kappa^{\mu\nu\lambda\rho}$ having dimension of inverse
of mass. In the QED extension with this additional vertex, radiative
corrections at first and second order of the LV coefficients were
presented in the literature. The dominant contributions are given
by\,\citep{Casana:2013nfx} 
\begin{equation}
{\cal L}_{eff}\supset\frac{me}{8\pi^{2}\epsilon}\kappa^{\mu\nu\lambda\rho}F_{\mu\nu}F_{\lambda\rho}+{\rm finite},
\end{equation}
which matches the minimal CPT-even $k_{F}$ term in the SME, contributing
to its renormalization. At the second order, besides a minimal $k_{F}$
term of the form $(k_{F})^{\mu\nu\lambda\rho}\propto\kappa^{\mu\nu\alpha\beta}\left(\kappa\right)_{\alpha\beta}^{\phantom{\alpha\beta}\lambda\rho}$,
one also will obtain the higher-derivative terms 
\begin{equation}
{\cal L}_{eff}\supset\frac{1}{\pi^{2}\epsilon}\left(C_{5}\kappa^{\mu\nu\alpha\beta}\kappa_{\alpha\beta}^{\phantom{\alpha\beta}\lambda\rho}F_{\mu\nu}\Box F_{\lambda\rho}+C_{6}\kappa^{\mu\nu\alpha\beta}\kappa_{\beta}^{\phantom{\beta}\gamma\lambda\rho}F_{\mu\nu}\partial_{\alpha}\partial_{\gamma}F_{\lambda\rho}\right)+{\rm finite},
\end{equation}
where $C_{5}$ and $C_{6}$ are dimensionless numerical constants.
In the two last expressions, finite parts, in the UV leading order,
reproduce the same tensorial structures as the pole parts.

A natural modification of the previous example consists in introducing
a pseudotensor coupling\,\citep{Araujo:2016hsn}, 
\begin{equation}
{\cal V}_{4}=-ig\kappa^{\mu\nu\lambda\rho}\thinspace\bar{\psi}\sigma_{\mu\nu}\gamma_{5}\psi\thinspace F_{\lambda\rho}=-\frac{1}{4}H_{F}^{\left(5\right)\mu\nu\alpha\beta}\thinspace\bar{\psi}\sigma_{\mu\nu}\psi\thinspace F_{\alpha\beta}\thinspace,
\end{equation}
where now $H_{F}^{\left(5\right)\mu\nu\alpha\beta}=-2g\kappa^{\mu\nu\lambda\rho}\epsilon_{\lambda\rho}^{\hphantom{\lambda\rho}\alpha\beta}$.
In this case, the resulting quantum corrections to the photon sector
we will involve a \textquotedbl twisted\textquotedbl{} tensor $\bar{F}^{\mu\nu}=\kappa^{\mu\nu\lambda\rho}F_{\lambda\rho}$,
together with the dual $\tilde{F}_{\mu\nu}=\frac{1}{2}\varepsilon_{\mu\nu\alpha\beta}F^{\alpha\beta}$.
Again, as in the previous case, we will have contributions involving
both second and higher derivatives, and they are divergent\,\citep{Carvalho:2018vtr}.
The explicit result reads 
\begin{align}
{\cal L}_{eff}\supset & \frac{1}{16\pi^{2}\epsilon}\left(8e\bar{F}^{\mu\nu}\tilde{F}_{\mu\nu}+8m^{2}\bar{F}^{\mu\nu}\tilde{F}_{\mu\nu}+8m^{2}\bar{F}^{\alpha\gamma}\partial_{\alpha}\partial_{\beta}\bar{F}_{\phantom{\beta}\gamma}^{\beta}+\right.\nonumber \\
 & \left.+2m^{2}\lambda_{2}^{2}\bar{F}^{\alpha\gamma}\partial^{\beta}\partial^{\lambda}\bar{F}^{\sigma\rho}\epsilon_{\alpha\gamma\beta\mu}\epsilon_{\sigma\rho\lambda}^{\phantom{\sigma\rho\lambda}\mu}\right)+{\rm finite}.
\end{align}
This result includes terms both of first and in second orders in the
LV coefficient $\kappa_{\mu\nu\lambda\rho}$. We note that some of
these contributions are CPT-odd despite the even order in derivatives,
because of the Levi-Civita symbol. The tensorial structure of the
finite parts is again analogous to that one of pole parts. Some consequences
of these generated terms have been studied in\,\citep{Carvalho:2018vtr}.

Another interaction considered in the literature, in order to generate
higher-derivatives contributions in the gauge sector, is based on
the Myers-Pospelov approach\,\citep{MP}. The idea is that additional
derivatives appear in the action being contracted to some constant
vector, which as a result prevents the arising of ghosts, and in the
Lorentz-invariant limit the higher derivatives disappear completely.
One may start by adding to the QED Lagrangian the following term,
\begin{equation}
{\cal V}_{5}=\frac{1}{M^{n-1}}\bar{\psi}\gamma_{5}\slashed{v}(v\cdot D)^{n}\psi\thinspace,\label{verthd}
\end{equation}
with $n\geq2$. Here $M$ is the energy scale supposed of the order
of the Planck mass. This operator has mass dimensions equal to $n+3$,
and have been studied in the case $n=2$\,\citep{CMR2}, where the
linear combination of the higher-derivative CFJ-like term\,(\ref{HDCFJ})
and the Myers-Pospelov term\,(\ref{MP}) was generated, both being
divergent. In principle, many other terms can be generated from the
couplings\,(\ref{verthd}), for example, it is natural to expect
that the aether term can arise at least for some values of $n$.

\section{Spinor-scalar LV couplings\label{sec:Spinor-scalar-LV-couplings}}

The spinor-scalar LV couplings are studied in a smaller number of
papers compared with the ones discussed so far, yet several interesting
results have been presented in the literature. For example, the Yukawa
potential was calculated in\,\citep{Altschul:2006jj} considering
LV just in the scalar sector, and in\,\citep{Altschul:2012xu} for
a LV modification of the spinor-scalar coupling of the form $\bar{\psi}G\psi\phi$,
with $G$ being a matrix of the form 
\begin{equation}
G=g+ig'\gamma_{5}+a^{\mu}\gamma_{\mu}+b^{\mu}\gamma_{5}\gamma_{\mu}+\frac{1}{2}L^{\mu\nu}\sigma_{\mu\nu}\thinspace.
\end{equation}
Also, the one loop renormalization of a general model including fermions
and scalars interacting including the aforementioned general LV Yukawa
coupling was worked out in\,\citep{Ferrero:2011yu}. 

Lorentz violating Yukawa couplings have also appeared in other studies
that looked into radiative corrections. The particular LV coupling
\begin{equation}
{\cal Y}_{1}=a_{\mu}\thinspace\bar{\psi}\gamma^{\mu}\psi\thinspace\phi\thinspace,\label{vertscal}
\end{equation}
has been used in\,\citep{aether} in order to generate the CPT-even
aether-like term for the scalar field, 
\begin{equation}
{\cal L}_{eff}\supset C_{7}\phi(a\cdot\partial)^{2}\phi\thinspace,
\end{equation}
where $C_{7}$ is a constant which diverges in four-dimensional space-time.
This contributions amounts to $\left(k_{c}^{(4)}\right)^{\mu\nu}=C_{7}a^{\mu}a^{\nu}$
in the SME conventions put forth in\,\citep{Edwards:2018lsn}. Actually,
the aether term for the scalar field is the simplest Lorentz-breaking
contribution for the scalar sector in the four-dimensional space-time,
and in\,\citep{CMR3}, it has been shown to arise also for the Lorentz-breaking
spinor-scalar theory with the usual Yukawa coupling, but with Myers-Pospelov-like
higher-derivative modified kinetic term for the spinor, being finite
in this case.

Another interesting coupling in this sector is the pseudoscalar one,
\begin{equation}
{\cal Y}_{2}=b_{\mu}\bar{\psi}\gamma^{\mu}\gamma_{5}\psi\thinspace\vartheta,\label{vertscal1}
\end{equation}
where $\vartheta(x)$ is a pseudoscalar field. This vertex has been
used in\,\citep{axion} as a part of a mechanism to generate the
photon-axion term (\ref{axion}), as discussed in the previous section. 

It is clear that, in principle, terms with more derivatives can be
generated in the scalar sector as well, by the same couplings above,
considering the derivative expansion of the two-point vertex function
of the scalar, where the finiteness of these terms will be guaranteed
by the renormalizability of the couplings (\ref{vertscal},\ref{vertscal1}).
In principle, these couplings can be used to generate the interaction
terms for the Lorentz-breaking Higgs sector, although these calculations
were not carried out up to now, at least in the approach we are considering.
It is easy to see that since these couplings are dimensionless, the
corresponding contributions to the vertices in the Higgs sector will
be logarithmically divergent. Finally, it is worth mentioning the
study of the Lorentz-breaking Higgs sector carried out for the scalar
QED in\,\citep{Scarp2013}, where the spontaneous symmetry breaking
is considered in detail, including one-loop quantum corrections.

\section{LV contributions in the spinor sector\label{sec:LV-spinor}}

The number of studies concerning the generation of LV corrections
in the fermion sector of the SME is much smaller than for the scalar
and especially the gauge sectors. In the context of LV theories arising
as effective field theories, it seems natural to start with some basic
Lorentz-breaking theory where the scalar and gauge fields are originally
coupled to spinor fields, which are then integrated out. This was
the basic mechanism considered in most of the works mentioned in the
previous sections. Nevertheless, the study of the spinor-dependent
LV contributions is an interesting task, as we discuss in this section.

At tree level, the LV extension of the spinor sector of the Standard
Model was first described in the seminal papers \citep{coll1,coll2},
with the restriction of minimal (renormalizable) operators. More recently,
the non-minimal fermionic sector with LV was described in\,\citep{KosMew},
presenting a general parametrization for the LV coefficients, as well
as discussing several aspects of these models such as dispersion relations,
exact Hamiltonian and eigenstates, together with some first numerical
estimations for these Lorentz-breaking parameters.

Regarding the quantum corrections, for the minimal sector, an exhaustive
study of the one-loop divergent contributions to the spinor sector
for the QED sector of the SME has been presented in\,\citep{KostPic},
where the full one-loop renormalization of this sector was studied.
In the non-minimal sector, one first result was that the CPT-odd term
\begin{equation}
{\cal S}_{1}=c^{\mu\nu}\bar{\psi}\gamma_{\mu}\partial_{\nu}\psi\thinspace,
\end{equation}
was shown to arise in the non-minimal extension of the QED developed
in\,\citep{aether} based on the non-minimal magnetic coupling (\ref{v1n}),
where $c^{\mu\nu}\sim b^{\mu}b^{\nu}$, the proportionality involving
a divergent constant that needs to be renormalized. The contribution
of the same structure was shown in\,\citep{Belich} to arise also
from the CPT-even coupling (\ref{casana}), where, however, a particular
form of the $\kappa_{\mu\nu\lambda\rho}$ tensor completely described
by one vector $u^{\mu}$ has been used. Again, this contribution diverges.
Also, in\,\citep{CMR3}, the Lorentz-breaking extension of the Yukawa
model with the extra term 
\begin{equation}
{\cal S}_{2}=\bar{\psi}(\alpha m+g(a\cdot\partial)^{2})\slashed{a}\psi
\end{equation}
$\alpha$ being a constant, has been considered and divergent quantum
corrections where shown to arise for the first term in ${\cal S}_{1}$.

Taking into account all this, one can expect that the number of open
problems in the spinor sector is still very large, and the most important
among them are studying of quantum contributions to the kinetic term
of the spinor field and to spinor-scalar (and similarly, spinor-vector)
couplings, for various Lorentz-breaking extensions of QED and Yukawa
model, both minimal and non-minimal ones.

\section{Lorentz violation in Gravity\label{sec:Gravity}}

The problem of quantum corrections with Lorentz-breaking extensions
in gravity is much more complicated than with extension of other field
theory models. The first reason for this is that general coordinate
invariance, being the essential symmetry of the Einstein gravity,
is known to play a double role, being at the same time the analogue
not only of the Lorentz symmetry, but also of a gauge symmetry. Therefore,
the breaking of the general coordinate invariance will be associated
with the breaking of gauge symmetry (for an extensive discussion of
possible implications of breaking the gauge symmetry in the linearized
gravity, see\,\citep{KostMewnew}). The second reason is that in
the case of the curved space-time, the constant vectors or tensors
used in the non gravitational SME to introduce preferred space-time
directions are almost useless since their covariant derivatives are
not necessary equal to zero, implying in a great increasing of the
number of new structures which involve covariant derivatives of these
\textquotedbl constant\textquotedbl{} tensors, see f.e. \citep{ShapNet}.
Third, it has been noted that backgrounds with explicit Lorentz violation
are not in general compatible with Riemannian geometry\,\citep{KosGra},
and there are hints that an alternate geometrical description of gravity,
based on Riemann-Finsler geometries, may be necessary to describe
explicit LV backgrounds\,\citep{Kostelecky:2011qz,Kostelecky:2012ac,Edwards:2018lsn}
in a consistent way. Besides these questions, one has the well-known
difficulty of quantum calculations in gravity arising from the fact
that the Einstein gravity is a highly nonlinear, and, moreover, non-renormalizable
theory. Apart from these theoretical questions, a strong phenomenological
program have been developed, connecting the gravitational SME with
the Post-Newtonian formalism\,\citep{Bailey:2006fd}, and setting
the grounds for looking for Lorentz violation in short-range gravity
experiments\,\citep{Bailey:2014bta,Shao:2016cjk,Kostelecky:2016uex},
gravitational Cerenkov radiation\,\citep{Kostelecky:2015dpa} and
gravitational waves\,\citep{Kostelecky:2016kfm,Kostelecky:2016kkn},
among others. 

One of the early models including Lorentz violation in gravity that
where studied was devoted to the four-dimensional Chern-Simons (CS)
modified gravity which, besides the usual Einstein-Hilbert Lagrangian,
includes the CS term\,\citep{JaPi} 
\begin{equation}
S_{CS}=\frac{1}{64\pi G}\int d^{4}x\vartheta{}^{*}RR\thinspace,\label{gravcs}
\end{equation}
where 
\begin{equation}
^{*}RR={}^{*}R_{\phantom{\sigma}\tau}^{\sigma\phantom{\tau}\mu\nu}R_{\phantom{\tau}\sigma\mu\nu}^{\tau}=\frac{1}{2}\epsilon^{\mu\nu\alpha\beta}R_{\phantom{\sigma}\tau}^{\sigma\phantom{\tau}\alpha\beta}R_{\phantom{\tau}\sigma\mu\nu}^{\tau}\thinspace.
\end{equation}
This term is a total derivative, just as the $\tilde{F}^{\mu\nu}F_{\mu\nu}$
in the electromagnetism, therefore, there is a natural analogy between
the CFJ term and the $4D$ gravitational CS term. In general, $S_{CS}$
breaks only the CPT symmetry, becoming Lorentz-violating for a special
choice of the $\vartheta$ in the form $\vartheta=k_{\mu}x^{\mu}$,
with $k_{\mu}$ being a constant axial vector. In\,\citep{JaPi},
the consistency of such a choice for $\vartheta$ has been proved
for the case of the Schwarzschild metric; in general, it is achieved
if the condition $^{*}RR=0$ is satisfied, and this indeed occurs
for a wide class of metrics with spherical or axial symmetry\,\citep{Grumiller}.
In terms of the spin connection $\omega_{\nu ab}$, this LV form of
the $4D$ gravitational CS model can be written as 
\begin{equation}
S_{CS}=\frac{1}{64\pi G}\int d^{4}x\thinspace\epsilon^{\mu\nu\lambda\rho}k_{\mu}\left(\partial_{\nu}\omega_{\lambda ab}\omega_{\rho}^{\phantom{\rho}ba}-\frac{2}{3}\omega_{\nu ab}\omega_{\lambda}^{\phantom{\lambda}bc}\omega_{\rho c}^{\phantom{\rho c}a}\right)\thinspace.\label{csg}
\end{equation}

While the study of classical aspects of the $4D$ CS modified gravity
is clearly an interesting problem (for a review, see\,\citep{Alexander}),
we are mainly interested in the perturbative generation of LV operators.
It is clear that the consideration of the weak gravity approximation
greatly simplifies the calculations. In the linear approximation,
equation\,\eqref{csg} can be written as
\begin{equation}
S_{CS}^{lin}=\frac{1}{256\pi G}\int d^{4}xk^{\lambda}h^{\nu\alpha}\epsilon_{\mu\nu\lambda\rho}\partial^{\rho}\left(\Box h_{\alpha}^{\mu}-\partial_{\alpha}\partial_{\gamma}h^{\gamma\mu}\right)\thinspace.\label{linform}
\end{equation}
The general formalism for LV in linearized gravity was developed in\,\citep{KostMewnew},
and in term of these conventions, the linearized form of the gravitational
CS term in the last equation corresponds to the dimension five, gauge-invariant
and CPT-odd coefficient
\begin{equation}
q^{\left(5\right)\mu\rho\epsilon_{1}v\epsilon_{2}\sigma\epsilon_{3}}=\frac{3}{4}k_{\lambda}\epsilon^{\mu\rho\epsilon_{1}\lambda}\left(\eta^{\epsilon_{2}\epsilon_{3}}\eta^{\nu\sigma}-\eta^{\epsilon_{2}\sigma}\eta^{\nu\epsilon_{3}}\right)\thinspace.
\end{equation}

The linearized $4D$ gravitational CS term has been generated at the
one-loop order for the first time in\,\citep{gravCS}, starting with
the following action of the spinor field in the curved space-time,
\begin{equation}
S=\int d^{4}x\thinspace\left(\frac{i}{2}ee_{a}^{\mu}\bar{\psi}\gamma^{a}\stackrel{\leftrightarrow}{D}_{\mu}\psi-ee_{a}^{\mu}\bar{\psi}b_{\mu}\gamma^{\mu}\gamma_{5}\psi+em\bar{\psi}\psi\right)\thinspace,\label{spingrav}
\end{equation}
where 
\[
D_{\mu}\psi=\left(\partial_{\mu}+\frac{1}{2}\omega_{\mu cd}\sigma^{cd}\right)\psi\thinspace,
\]
$e_{a}^{\mu}$ is the vielbein, $e$ its determinant, and $b_{\mu}$
is a constant axial vector. As a result, despite the corresponding
Feynman diagrams being superficially divergent, the result of the
calculations turns out to be finite due to the gauge symmetry, reproducing
Eq.\,(\ref{linform}) with $k^{\lambda}=\frac{1}{48\pi^{2}}b^{\lambda}$.
Similarly to the case of the quantum generation of the CFJ term\,\citep{JackAmb},
the corresponding mechanism for the generation of the $4D$ gravitational
CS term was also shown to be ambiguous\,\citep{ptime,ambgrav}. 

The action\,\eqref{spingrav} was also considered in\,\citep{lingrav},
in order to generate the LV term 
\begin{equation}
{\cal L}_{eff}\supset C_{g}m^{2}\epsilon^{\mu\nu\lambda\rho}b_{\rho}h_{\nu\sigma}\partial_{\lambda}h_{\mu}^{\sigma}\thinspace,
\end{equation}
corresponding, in the SME notation of\,\citep{KostMewnew}, to the
LV coefficient
\begin{equation}
q^{\left(3,3\right)\mu\rho\epsilon_{1}\nu\sigma}=4C_{g}m^{2}\epsilon^{\mu\rho\epsilon_{1}\nu\sigma}\eta^{\nu\sigma}\thinspace,
\end{equation}
which was argued to be related to a non commutative geometric setup.
However, the constant $C_{g}$ is divergent, hence, for a consistent
renormalization it must be introduced from the very beginning. Besides
of this, this term breaks gauge invariance in general. A particular
setup of wave propagation was discussed in\,\citep{lingrav}, exhibiting
vacuum birefringence, but the question of the breaking of gauge invariance
was not solved in general.

Here, it should be mentioned that various Lorentz-breaking terms in
the gravitational sector like $\phi^{\mu\nu}R_{\mu\nu}$, $\phi^{\mu\nu\alpha\beta}R_{\mu\nu\alpha\beta}$,
with $\phi^{\mu\nu}$ and $\phi^{\mu\nu\alpha\beta}$ being constructed
from covariant derivatives of constant tensors present in Eq.\,(\ref{genrenmod})
have been generated in\,\citep{ShapNet} for a theory represented
itself as a general model like (\ref{genrenmod}) embedded into a
curved space-time. However, all these terms, and even terms of higher
orders in the curvature tensor, are essentially divergent.

As noted before, the generic incompatibility of explicit Lorentz violation
with the geometric interpretation of General Relativity means that
scenarios of LV being spontaneously broken are particularly interesting.
This idea was proposed already in \citep{KosSam} and can be naturally
promoted to a curved space-time. One way to do it is based on the
Einstein-aether model (see f.e. \citep{JaMat}) whose action looks
like 
\begin{equation}
S=-\frac{1}{16\pi G}\int d^{4}x\sqrt{|g|}\left(R+K_{\phantom{\alpha\beta}\mu\nu}^{\alpha\beta}\nabla_{\alpha}u^{\mu}\nabla_{\beta}u^{\nu}+\lambda(u^{\mu}u_{\mu}-1)\right)\thinspace,
\end{equation}
where 
\begin{equation}
K_{\phantom{\alpha\beta}\mu\nu}^{\alpha\beta}=c_{1}g^{\alpha\beta}g_{\mu\nu}+c_{2}\delta_{\mu}^{\alpha}\delta_{\nu}^{\beta}+c_{3}\delta_{\nu}^{\alpha}\delta_{\mu}^{\beta}+c_{4}u^{\alpha}u^{\beta}g_{\mu\nu},
\end{equation}
with $c_{1},c_{2},c_{3},c_{4}$ are some dimensionless constants,
the $\lambda$ is a Lagrange multiplier used to implement the constraint
$u^{\mu}u_{\mu}-1=0$. Further development of this concept was presented
by the bumblebee model \citep{Bertolami}, where, instead of the constraint
on the vector field $B_{\mu}$, called now the bumblebee field, a
potential for $B_{\mu}$ is introduced, whose various vacua generate
different preferred space-time directions (nevertheless, it should
be noted that the vacua in general are no more represented by constant
vectors). Moreover, the vector field becomes a dynamical field in
its own nature. The corresponding action is 
\begin{equation}
S=-\frac{1}{16\pi G}\int d^{4}x\sqrt{|g|}\left(R+\xi B^{\mu}B^{\nu}R_{\mu\nu}-\frac{1}{4}B_{\mu\nu}B^{\mu\nu}-V\left(B^{\mu}B_{\mu}-b^{2}\right)\right)\thinspace,
\end{equation}
with $\xi$ and $b^{2}$ being constants, and the $B_{\mu\nu}$ the
stress tensor for the bumblebee field. In this case, the quantum dynamics
of the bumblebee field could be considered, and in principle one could
expect some generalization of the result from \citep{Gomes:2007mq}
for a curved space-time. However, up to now, there are no examples
of loop calculations for a bumblebee theory on a curved background.

Without trying to exhaust the topic, we conclude that the problem
of Lorentz violation in gravity is still a very open one, where one
expects to see new developments, both on the many conceptual issues,
as well as from the phenomenological standpoint. In particular, the
matter of radiative corrections is still quite unexplored.

\section{Conclusions\label{sec:Conclusions}}

There have been extensive activity in the search for possible Lorentz
violation in the last decades, which have resulted in a solid experimental
programme\,\citep{datatables}, as well as in a deep understanding
of the theoretical questions involved in incorporating CPT and/or
Lorentz breaking in the context of effective field theory. From the
theoretical viewpoint, the question of quantum corrections and its
effects when LV operators are considered is one of the most studied,
since the seminal papers\,\citep{coll2,JK}. In this work, we revisit
this question, filling many gaps present in the literature, when results
were not written in the standard SME notation, or their phenomenological
implications not fully addressed. In many instances, the generated
operators are \textquotedbl finite but undetermined\textquotedbl ,
or divergent, however there are examples where finite, well defined
corrections can be shown to exist, and these may even lead to improved
experimental bounds on some LV coefficients. The most natural scenario
for these studies involves the calculation of radiative corrections
generated by fermion loops involving the minimal LV coefficients of
the QED sector of the SME. Since this represents a small set of the
possible coefficients for LV, most of these cases have been already
considered in the literature, at least in the leading order. In this
work, however, we show that higher order corrections may still provide
interesting results, such as the example contained in Eq.\,\eqref{aether},
and these have not been so systematically addressed. Also, when this
idea is extended to the non minimal SME, despite a few specific setups
that where devised to generate specific \textquotedbl notorious\textquotedbl{}
LV operators, systematic studies are still missing, in particular
in sectors others than the extended QED. We conclude that still there
is space for studying quantum corrections in LV theories, and that
this program may even help the experimental task of constraining Lorentz
violation via different experiments.

\medskip{}

\textbf{Acknowledgments.} The authors would like to thank V. A. Kostelecky
for discussions and interesting insights that helped us to improve
our paper. This work was partially supported by Conselho Nacional
de Desenvolvimento Científico e Tecnológico (CNPq) and Fundação de
Amparo à Pesquisa do Estado de São Paulo (FAPESP), via the following
grants: CNPq 304134/2017-1 and FAPESP 2017/13767-9 (AFF), CNPq 303783/2015-0
(AYP).

\appendix

\section{Summary Table}

In this Appendix, we present a table summarizing the results concerning
the generation of effective operators in the photon sector of the
SME, originating from the LV operators in the spinor sector, as a
summary of the results discussed in Sections\,\ref{sec:minimalQED}
and\,\ref{sec:nonminimalQED}. In formulating these tables, it is
assumed that the usual vertex $\sim e\bar{\psi}\gamma^{\mu}\psi A_{\mu}$
can be combined with each of the Lorentz-violating vertices involving
the gauge field. For the $d^{\mu\nu}$ coefficient, the form presented
in the table is derived from covariance and symmetry arguments, since
no explicit results are available. For simplicity, non-Abelian generalizations
are not included, but they are cited in the main text. Also, all Lorentz
invariant contributions that are generated are omitted in the table,
exception made for the case involving the axion field $\vartheta$,
which is also the only one where the generation mechanism involves
two different LV couplings.

\begin{table}[H]
\centering{}\caption{Summary table of the generation of Lorentz-breaking operators in the
photon sector of the SME. Notice that CPT-odd coefficients can generate
CPT-even corrections at second order. Also, the axion case (pseudo-scalar
field $\vartheta$) is not written in SME notation since the induced
correction is actually Lorentz invariant, so it is not included in
the traceless coefficient $k_{F}$. The columns D and A indicate whether
the diagrams generating the given operator are divergent or ambiguous,
respectively. }
\vspace*{1mm}
 \label{tab:table1} %
\begin{tabular}{|l|c|c|c|c|r|}
\hline 
LV operator  & CPT & Generated term & D & A & References\tabularnewline
\hline 
$a_{\mu}$  & O & 0 &  &  & \citep{Scarp3}\tabularnewline
\hline 
$b_{\mu}$  & O & $(k_{AF}^{(4)})_{\mu}\sim b_{\mu}$ &  & X & \citep{coll2,Coleman:1998ti,Chung:1998jv,Andrianov:1998ay,JK}\tabularnewline
 &  & $(\hat{k}_{AF}^{\left(5\right)})_{\kappa}^{\hphantom{\kappa}\alpha\beta}\sim b_{\kappa}\eta^{\alpha\beta}$ &  &  & \citep{TMariz2}\tabularnewline
 &  & $(k_{F}^{(4)})^{\mu\nu\alpha\beta}\sim b^{\mu}b^{\alpha}\eta^{\nu\beta}+\cdots$ &  &  & \citep{Bonneau,aether2}\tabularnewline
\hline 
$c^{\mu\nu}$  & E & $(k_{F}^{(4)})_{\mu\nu\rho\sigma}\sim c_{\mu\rho}\eta_{\nu\sigma}+\cdots$ & X &  & \citep{Maluf}\tabularnewline
\hline 
$d^{\mu\nu}$  & E & $(k_{F}^{(4)})^{\mu\nu\rho\sigma}\sim d^{\mu\rho}d^{\nu\sigma}-d^{\mu\sigma}d^{\nu\rho}$ & X &  & \citep{Scarp3}\tabularnewline
\hline 
$e^{\mu}$  & O & $(k_{F}^{(4)})^{\mu\nu\rho\sigma}\sim\eta^{\rho\mu}e^{\nu}e^{\sigma}+\cdots$ & X &  & \citep{Scarp3}\tabularnewline
\hline 
$f^{\mu}$  & O & $(k_{F}^{(4)})^{\mu\nu\rho\sigma}\sim\eta^{\rho\mu}f^{\nu}f^{\sigma}+\cdots$ & X &  & \citep{Scarp3}\tabularnewline
\hline 
$g^{\mu\nu\lambda}$  & O & $(\hat{k}_{AF}^{\left(5\right)})_{\kappa}^{\hphantom{\kappa}\alpha\beta}\sim\epsilon_{\kappa\lambda\mu\nu}\left(g^{\mu\nu\alpha}\eta^{\rho\beta}+\cdots\right)$ &  &  & \citep{MarizHD}\tabularnewline
\hline 
$(a_{F}^{(5)})^{\alpha\beta\gamma}=2g\thinspace\epsilon^{\rho\alpha\beta\gamma}u_{\rho}$ & O & $(k_{F}^{(4)})^{\mu\nu\alpha\beta}\sim b^{\mu}b^{\alpha}\eta^{\nu\beta}+\cdots$ &  & X & \citep{aether}\tabularnewline
\hline 
$(a_{F}^{(5)})^{\alpha\beta\gamma}=-\left(u^{\beta}\eta^{\gamma\alpha}-u^{\alpha}\eta^{\gamma\beta}\right)$ & O & ${\cal L}_{eff}\supset\epsilon^{\mu\nu\alpha\beta}b_{\alpha}u^{\rho}\thinspace\vartheta F_{\mu\nu}F_{\rho\beta}$ &  & X & \citep{axion}\tabularnewline
$\bar{\psi}\gamma_{5}\gamma^{\mu}\psi\thinspace b_{\mu}\thinspace\vartheta$  & O &  &  &  & \tabularnewline
\hline 
$H_{F}^{(5)\mu\nu\alpha\beta}=-\frac{1}{2}\kappa^{\mu\nu\lambda\rho}$ & E & $(k_{F}^{(4)})_{\mu\nu\lambda\rho}\sim k_{\mu\nu\lambda\rho}$ & X &  & \citep{Casana:2013nfx}\tabularnewline
\hline 
$H_{F}^{\left(5\right)\mu\nu\alpha\beta}=-2g\kappa^{\mu\nu\lambda\rho}\epsilon_{\lambda\rho}^{\hphantom{\lambda\rho}\alpha\beta}$ & E & $(k_{F}^{(4)})_{\mu\nu\lambda\rho}\sim k_{\mu\nu\alpha\beta}\epsilon_{\hphantom{\alpha\beta}\lambda\rho}^{\alpha\beta}$ & X &  & \citep{Araujo:2016hsn}\tabularnewline
\hline 
\end{tabular}
\end{table}

\end{document}